# Variations on the Newcomb-Benford Law

Subhash Kak

**Abstract.**
The Newcomb-Benford Law, which is also called the first digit phenomenon, has applications in diverse phenomena ranging from social and computer networks, engineering systems, natural sciences, and accounting. In forensics, it has been used to determine intrusion in a computer server based on the measured expectations of first digits of time varying values of data, and to check whether the information in a data base has been tampered with. There are slight deviations from the law in certain natural data, as in fundamental physical constants, and here we propose a more general bin distribution of which the Newcomb-Benford Law is a special case so that it can be used to provide a better fit to such data, and also open the door to a mathematical examination of the origins of such deviations.

**Introduction**
According to the Newcomb-Benford law on first digits (which will be shortened to NB Law and NB distribution in the remainder of the paper), the probabilities of first digits are according to the logarithmic distribution related to the digit values. It is interpreted in different ways ranging from scale invariance to a property that emerges from underlying random counting process [1][2]. The observation on the first digit phenomenon was first made by Newcomb in 1881 and rediscovered by Benford in 1938; the law is also called Benford's Law.

The NB distribution has been used in a variety of applications ranging from social and computer networks, forensics, and accounting [3],[4]. Sambridge et al [5] report its applicability also to modern observational data drawn from a variety of natural physical, engineering, geological and other fields. Clearly this distribution is also important in the study of cryptographic keys obtained from natural data, as in certain protocols.

If a counting process is uniformly distributed over the range {1, ..., $S$}, with random values of $S$, then the sum of a large number of these will satisfy the NB Law [4], where the leading digit $d$ ($d \in \{1, ..., 9\}$) for number to the base 10 occurs with probability as a logarithmic function:

$$P(d) = \log_{10}\left(1 + \frac{1}{d}\right) \tag{1}$$

In general, the distribution of first digits for numbers in an arbitrary base *r*, *r* ≥ 2, is:

$$P(d) = \log_r\left(1 + \frac{1}{d}\right) \tag{2}$$

For a number consisting of several digits, the same law applies with *d* replaced by the number, which shows that the digits are dependent. This also makes it possible to obtain the probability distribution for the second and subsequent digits. This dependence amongst digits decreases as the distance between them increases. But since some data (to be shown next) has deviations from the normative NB distribution, it opens up the question of the origins of the dependence.

As is to be expected, the probability values get progressively smaller beyond the first digit. Table 1 gives the predicted frequencies for first to fourth digits and as we can see the probabilities tend to become uniform as we go to higher digit frequencies.

Table 1. First to fourth digit frequencies

|  | 0 | 1 | 2 | 3 | 4 | 5 | 6 | 7 | 8 | 9 |
|---|---|---|---|---|---|---|---|---|---|---|
| 1st digit | - | 0.3010 | 0.1761 | 0.1249 | 0.0969 | 0.0792 | 0.0669 | 0.0580 | 0.0512 | 0.0458 |
| 2nd digit | 0.1197 | 0.1139 | 0.1088 | 0.1043 | 0.1003 | 0.0967 | 0.0934 | 0.0904 | 0.0876 | 0.0850 |
| 3rd digit | 0.1018 | 0.1014 | 0.1010 | 0.1006 | 0.1002 | 0.0998 | 0.0994 | 0.0990 | 0.0986 | 0.0983 |
| 4th digit | 0.1002 | 0.1001 | 0.1001 | 0.1001 | 0.1000 | 0.1000 | 0.0999 | 0.0999 | 0.0999 | 0.0998 |

NB's law is scale invariant [3]. In other words, if numbers in the data set are rescaled to another base, the probabilities will be adjusted appropriately for the new base. Thus if numbers are represented to base 4, the first digit probabilities will be P(1)=log$_4$(2)=0.5; P(2)=log$_4$(3/2)=0.292; and P(3)=log$_4$(4/3)=0.2075. In other words, half the random numbers to base 4 will begin with the digit 1.

The scale invariance of the NB distribution may be seen by comparing the values for numbers in different bases. The probability of a specific digit d < $r_1$, $r_2$, which are the two bases under consideration, is

$$\frac{P_{r_1}(d)}{P_{r_2}(d)} = \frac{\log_{r_1}(\frac{d+1}{d})}{\log_{r_2}(\frac{d+1}{d})} \tag{3}$$

Solving this, we get that

$$\frac{P_{r_1}(d)}{P_{r_2}(d)} = \frac{\log(r_2)}{\log(r_1)} \tag{4}$$

Thus the digit frequencies for 1, 2, and 3 in base 4 are log(4)/log(10) of the values for the same digits to base 10 that are given in Table 1.

In recent work, deviations from NB's Law were used to determine fake followers in social networks [6] and, by examining the incoming and outgoing data, whether there is an intrusion in a computer server [7]. These applications are in the same spirit as using the law to determine if a database has been tampered with [8] or its application to the study of hydrological and related data [9],[10]. It is also of potential importance in optimizing arithmetic processors of non-binary systems and understanding generalized numerical distributions (of which [11] is an example) and in generating random sequences that satisfy specific constraints in cryptographic systems, or perhaps create sequences that conform to the statistics associated with natural data [12]-[15]. The relationship of the NB distribution to other common distributions is in [16],[17]. To test whether data departs from the NB distribution, it is recommended [18] that the sample sizes be at least 1000.

We see that the observational data from natural phenomena broadly conforms to the expectations of the NB Law but there are differences that can be as large as nearly 4% for specific digit frequencies [5]. In order to address the question of these deviations, we propose a general bin distribution of which the NB Law is a special case.

**Derivation of the NB distribution**
Adhikari and Sarkar showed [19] that a set of equally distributed random numbers, when raised to higher and higher powers, will tend to NB's law. Table 2 shows that starting with 60,000 random numbers equally distributed with first digits that are 1 through 9, by the time they are raised to the 8[th] power, the probabilities have shifted quite close to those of NB's Law.

The NB distribution may be seen as the limiting distribution of successive exponentiations of a set of uniformly distributed numbers. And once this is arrived at, further operations leave it unchanged.

Table 2. Equally distributed 60,000 random numbers raised to higher powers [19]

| digits | power | | | | | | | |
|---|---|---|---|---|---|---|---|---|
| | 1 | 2 | 3 | 4 | 5 | 6 | 7 | 8 |
| 1 | 6,765 | 11,528 | 13,610 | 14,786 | 15,310 | 15,636 | 15,901 | 16,360 |
| 2 | 6,599 | 8,745 | 9,545 | 9,741 | 9,897 | 10,042 | 10,362 | 10,357 |
| 3 | 6,618 | 7,613 | 7,536 | 7,486 | 7,583 | 7,592 | 7,592 | 7,382 |
| 4 | 6,655 | 6,479 | 6,255 | 6,281 | 6,164 | 6,036 | 6,034 | 5,975 |
| 5 | 6,683 | 5,889 | 5,576 | 5,263 | 5,275 | 5,259 | 5,094 | 5,011 |
| 6 | 6,697 | 5,464 | 4,862 | 4,714 | 4,648 | 4,561 | 4,349 | 4,380 |
| 7 | 6,631 | 4,946 | 4,625 | 4,207 | 4,014 | 4,017 | 3,990 | 4,003 |
| 8 | 6,665 | 4,755 | 4,174 | 3,956 | 3,687 | 3,589 | 3,459 | 3,449 |
| 9 | 6,687 | 4,580 | 3,817 | 3,566 | 3,422 | 3,268 | 3,219 | 3,083 |

It was shown [19] that the logarithmic distribution arises if a large number of equally distributed random numbers are multiplied. It was further shown that if X has log distribution of the most significant digit, so do CX and 1/X, where C is any constant.

One may view such operations as successive convolutions of the corresponding probability density functions where the results are folded back to the original range.

More conveniently, let us consider a one-to-one map over (0,1). If a random variable Y is uniformly distributed over (0,1), then the probability density function of the random variable $X=Y^n$ is obtained by solving

$$f_X(x)dx = f_Y(y)dy \qquad (5)$$

Since $\frac{dx}{dy} = ny^{n-1}$ and $y = x^{1/n}$, this yields,

$$f_X(x) = \frac{1}{nx^{\frac{n-1}{n}}}, \qquad \text{with } x \text{ over } (0,1) \qquad (6)$$

As we can see, $\frac{1}{n}\int x^{(-n+1)/n} dx = \sqrt[n]{x}$, and therefore the total probability over (0,1) is 1.

The distribution of the first digit 1 can be computed by integrated f(x) over the range 0.1 to 0.2 and then 0.001 to 0.002, and so on over smaller and smaller ranges, and likewise for other digits as shown in Figure 1.

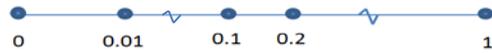

Figure 1. The recurrence of significant digit 1

Thus P(1) may be expressed as

$$P(1) = \int_{0.1}^{0.2} \sqrt[n]{x}\, dx + \int_{0.01}^{0.02} \sqrt[n]{x}\, dx + \int_{0.001}^{0.002} \sqrt[n]{x}\, dx + \cdots \qquad (7)$$

In the limit, as $n \to \infty$, the expression (7) reduces to (1) using the identity $\lim_{n\to\infty}(\frac{10}{x} - \frac{10}{x+1})^{1/n} = \left(\frac{10}{x(x+1)}\right)^{1/n} = 1$, as shown in [19].

**Entropy value with respect to radix**

If the leading digits had the same probability, the entropy associated with the r-1 digits will be $log(r-1)$. Normalizing it by dividing by the entropy of the r digits of the number system, we get the value $\frac{\ln(r-1)}{\ln(r)}$.

Figure 2 gives a chart of the entropy values of the first digit distribution with respect to radixes ranges from 3 to 10.

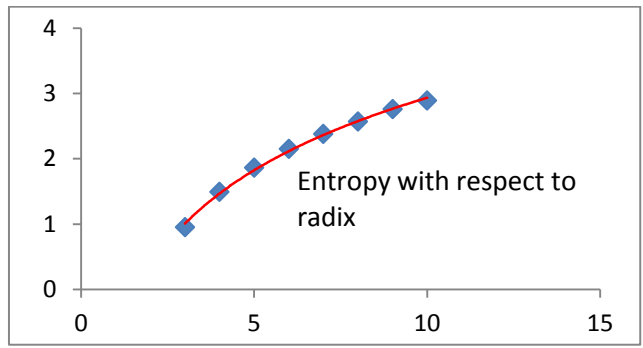

Figure 2. Entropy in bits with respect to radix from 3 to 10

When divided by the corresponding probability information associated with the radix, we get the following table:

Table 3. Entropy values of the first digits with respect to bases 3 to 10

| Radix | 3 | 4 | 5 | 6 | 7 | 8 | 9 | 10 |
|---|---|---|---|---|---|---|---|---|
| Entropy in bits | 0.95 | 1.49 | 1.86 | 2.15 | 2.38 | 2.57 | 2.76 | 2.89 |
| Entropy bits/ln(radix) | 0.86 | 1.07 | 1.16 | 1.20 | 1.22 | 1.24 | 1.26 | 1.26 |

The last row provides a relative comparison since bits are being divided by nats.

The first digit phenomenon presents another perspective on the relative efficiency of different bases where it is known that base-3 is the most efficient. Given that one wishes to represent numbers from 1 to N, the efficiency of the radix r is given by

$$E(r, N) \approx r \log_r N = r \frac{\ln N}{\ln r} \qquad (8)$$

Since 2/ln(2) ≈ 2.89 and 3/ln(3) ≈ 2.73, it follows that 3 is the integer base with the most superior radix economy [20].

**Frequency data from natural contexts**

We present observational data from a variety of natural phenomena in Table 4 which is taken from [5]. As one can see the data appears to largely conform to the frequencies expected from the NB's Law but there are small deviations. One can assume that these deviations are a consequence of the dependencies that exist in the digits and the relationship could offer insight into the nature of the data. It is conceivable that the deviations might offer clues on the underlying physical process.

Table 4. First digit data from a variety of natural phenomena, adapted from [5]

| | First Digit Frequencies | | | | | | | | |
|---|---|---|---|---|---|---|---|---|---|
| | 1 | 2 | 3 | 4 | 5 | 6 | 7 | 8 | 9 |
| NB's Law | 30.1 | 17.6 | 12.5 | 9.69 | 7.92 | 6.69 | 5.80 | 5.12 | 4.58 |
| Fund. Phys. constants | 34.0 | 18.4 | 9.2 | 8.28 | 8.58 | 7.36 | 3.37 | 5.21 | 5.52 |
| Geomagnetic field | 28.9 | 17.7 | 13.3 | 9.4 | 8.1 | 6.9 | 6.1 | 5.1 | 4.5 |
| Geomagnetic Reversals | 32.3 | 19.4 | 13.9 | 11.8 | 5.3 | 4.3 | 3.2 | 5.4 | 4.3 |
| Fermi Space Teles. fluxes | 30.3 | 17.9 | 13.0 | 9.9 | 7.6 | 6.96 | 5.23 | 5.23 | 2.72 |
| Pulsars rotation freq | 33.9 | 20.7 | 12.7 | 7.6 | 5.3 | 5.0 | 4.94 | 4.67 | 4.88 |

We see from Table 4, that in the second row there are over 3% departure from the expected values for digits 1 and 3. Likewise, in the fourth row, there are over 2% departure from the expected values for $2^{nd}$, $4^{th}$, $5^{th}$, $6^{th}$, and $7^{th}$ digits.

We would like to have a related distribution that can be made to fit this data. With this objective in mind, we consider the following general bin distribution which is developed in analogy with the NB distribution.

**A general bin distribution**

Consider *r-1* bins (1, 2, 3, ... *r-1*) in which objects are being placed randomly by different agents but in sequence. Let us represent *r* as follows:

$$r = \prod_{i=1}^{r-1} a_i \tag{9}$$

We can postulate a probability distribution for finding a specific object in one of the *r-1* bins by using the fact that

$$log_r r = 1 = log_r(\prod_{i=1}^{r-1} a_i) = \sum_{i=1}^{r-1} log_r(a_i) \tag{10}$$

This makes the probability, P(*i*), for bin *i*, to equal

$$P(i) = log_r a_i \tag{11}$$

This is the same as the NB distribution if

$$a_i = 1 + 1/i \tag{12}$$

In other words, the NB distribution factors are given as in the table below:

Table 5. NB distribution factors

| 2 | 3/2 | 4/3 | 5/4 | 6/5 | 7/6 | 8/7 | 9/8 | 10/9 |
|---|-----|-----|-----|-----|-----|-----|-----|------|

As we go from left to right, increasing terms in succession, the product ranges from 2, 3, and so on to 10, which is the base to the logarithmic function associated with the probabilities.

Now, consider the $a_i$ to be given by the following values:

$$a_1 = 2^i, a_2 = \frac{3^j}{2^i}, a_3 = \frac{4^k}{3^j}, \ldots, a_{r-1} = \frac{r}{(r-1)^s} \tag{13}$$

This will constitute a valid distribution so long as none of the probability values for the chosen exponentials exceeds 1. In other words, it requires that all relevant

$$\frac{(n+1)^{i+1}}{n^i} \leq 1 \tag{14}$$

The values of several of these exponents may be 1.

We will represent this distribution by the symbol P($i, j, k, \ldots$) based on the values of these numbers. The NB distribution corresponds to P(1,1,1, …).

The distribution P($i, i, i,\ldots$) has probabilities of the first digits that are $i$ times the values for the digits 1,…, $r$-2 and the correspondingly compensated value for the digit $r$-1.

Let's develop specific values of this distribution for specific data in Table 4. We first consider the second row on the fundamental physics constants.

To make probability of digit 1 to equal 0.34 of the Table, $log_{10}(2^i) = 0.34$.

A simple calculation shows that $i$=1.13.

To determine $j$, we solve $log_{10}(3^j/2^i) = 0.184$, and this gives us $j$=1.12.

To determine $k$, $log_{10}(4^k/3^j) = 0.092$, and $k$ = 1.04; the additional exponents turn out to be 2.95, 2.66, 2.39, 2.2, 0.98, and 0.945.

Thus the distribution of the first digits of theoretical physics constants of Table 4 corresponds to

$$P(1.13, 1.12, 1.04, 2.95, 2.66, 2.39, 2.2, 0.98, 0.945).$$

If only two values deviate from the NB frequencies, say between the third and the fourth digits, one can use a slight variation on Table 3, to account for it. The $4^i$ in the numerator of the third cell cancels out with the corresponding $4^i$ in the denominator of the 4$^{th}$ cell.

Table 4. Deviation for 3$^{rd}$ and 4$^{th}$ digits

| 2 | 3/2 | $4^i$/3 | 5/$4^i$ | 6/5 | 7/6 | 8/7 | 9/8 | 10/9 |
|---|---|---|---|---|---|---|---|---|

More complex tradeoffs between the frequency values may likewise be described.

**Conclusions**

There are slight variations from the law in certain natural data and this paper proposed a more general bin distribution of which the NB Law is a special case so that it may be used to provide a better fit to such data. There might very well be underlying physical processes that lead to the deviations from the expected values which the generalized NB distribution may help describe.